\documentclass[conference]{IEEEtran}
\IEEEoverridecommandlockouts

\usepackage{amsmath,amssymb,amsfonts}
\usepackage{graphicx}
\usepackage{fancyhdr}
\usepackage{float}
\usepackage{booktabs}
\usepackage{url}
\usepackage{xcolor}

\fancypagestyle{plain}{
    \fancyhf{}
    \fancyhead[L]{\footnotesize PREPRINT 2026}
    \fancyhead[R]{\footnotesize \thepage}
    
}

\begin{document}

\title{Desertification Forecasting in Semi-Arid Regions: Using Deep Learning to Probe the Observability Limits of Land Degradation Dynamics}

\author{
    \IEEEauthorblockN{Filipe Moura Da Silva}
    \IEEEauthorblockA{\textit{EEEP Avelino Magalhães} \\
    Tabuleiro do Norte, Ceará, Brazil \\
    Email: filipe.silva50@aluno.ce.gov.br}
    \and
    \IEEEauthorblockN{Cláubio Landney Lima Bandeira}
    \IEEEauthorblockA{\textit{Universidade Federal do Ceará} \\
    Russas, Ceará, Brazil}
    ORCID: 0009-0009-4220-3061
}

\maketitle
\thispagestyle{plain}
\pagestyle{plain}

\begin{abstract}
Forecasting desertification has received far less attention than mapping it. We use a convolutional LSTM (ConvLSTM) trained on ten annual Earth Observation (EO) variables (2000--2023) over the semi-arid state of Cear\'{a}, Brazil, on a 256$\times$206 grid at roughly 2.2 km resolution, to measure how much of the year-to-year evolution of a MapBiomas-derived degradation index (IDV) can be predicted from satellite data. The network predicts a bounded change on top of the last observed state, is trained in two stages (single-step, then recursive rollout) and is evaluated on a contiguous spatial and temporal holdout. State-wide mean IDV rose from 0.238 to 0.350 between 2000 and 2023, about 0.005 per year, but cell-level change is dominated by year-to-year jumps, and a gradient-boosting model given all channels explains only 4.7\% of its variance. Against this ceiling, the ConvLSTM lowers RMSE relative to persistence by 1.6\% one year ahead and 4.4\% four years ahead in a held-out southern band ($p < 0.001$), beats persistence with a constant drift and reproduces 53--69\% of the regional drift. Drift-aware baselines only help from the third year onward, by up to 4.4\% in RMSE at six years. An earlier formulation that redrew the whole field each year predicted changes unrelated to the observed ones (correlation $-0.06$) and produced a spurious decline when run recursively. Channel knockouts suggest that elevation and EVI degrade the forecast, while slope and aspect have no measurable effect. Monte Carlo dropout and stochastic scenario maps convey the uncertainty that the expected-value forecast hides. Under climatological forcing, mean IDV rises from 0.350 in 2023 to 0.370 in 2030. EO-based desertification forecasting is therefore reliable for the regional trend, not for individual cells.
\end{abstract}


\section*{Highlights}
\begin{itemize}
    \item Mean degradation in Cear\'{a} rose by about 0.005 IDV per year from 2000 to 2023.
    \item Less than 5\% of next-year change at 2.2~km is predictable from ten EO variables.
    \item A residual ConvLSTM beats persistence by 1--4\%, close to that ceiling.
    \item Redrawing the full field each year turns model error into spurious change.
\item Stochastic scenario maps reproduce realistic shares of unchanged, jumping and saturated cells, which the expected-value forecast cannot show.
\end{itemize}

\vfill\newpage
\section{Introduction}

A sustainable world depends on environmental monitoring systems able to read the present state of terrestrial ecosystems accurately and to anticipate where those ecosystems are heading under mounting climatic and anthropogenic pressure. Earth Observation (EO) is one of the primary scientific infrastructures behind that effort, supplying measurements of the land surface through satellite remote sensing \cite{zhao2025advancing}. EO archives have grown over the last two decades, and together with advances in artificial intelligence that growth has pushed the field toward predictive environmental intelligence \cite{janowicz2025geofm}. The most recent work goes further. Large pretrained foundation models now learn reusable geospatial representations from heterogeneous, multimodal EO sources, placing multispectral imagery and climatic products in a single learned representation \cite{janowicz2025geofm,zhou2025advances}. These models transfer more easily across remote sensing tasks and reduce the cost of adapting to new ones, which has made them a central computational framework for EO with clear relevance to climate adaptation and the SDGs, particularly SDG 13 (Climate Action) and SDG 15 (Life on Land) \cite{zhao2025advancing,un2015agenda,zhu2026foundations}.

Most of this progress has gone toward image interpretation and land-cover classification. General-purpose forecasting, and desertification forecasting in particular, remains comparatively underexplored \cite{janowicz2025geofm,zhou2025advances,zhu2026foundations}. Forecasting asks a different question from classification: which dynamical mechanisms govern how the state evolves under climatic, ecological and anthropogenic conditions that keep shifting \cite{zhu2026foundations}? It is a problem of temporal reasoning under uncertainty, and the influence of partially observed processes grows as the horizon lengthens \cite{zhu2026foundations}. Methods that represent these dynamics, and that locate the point at which partial observability starts to dominate forecast error, are one of the open directions in Geospatial Artificial Intelligence \cite{janowicz2025geofm,zhu2026foundations}.

\begin{figure}[!t]
    \centering
    \includegraphics[width=\columnwidth]{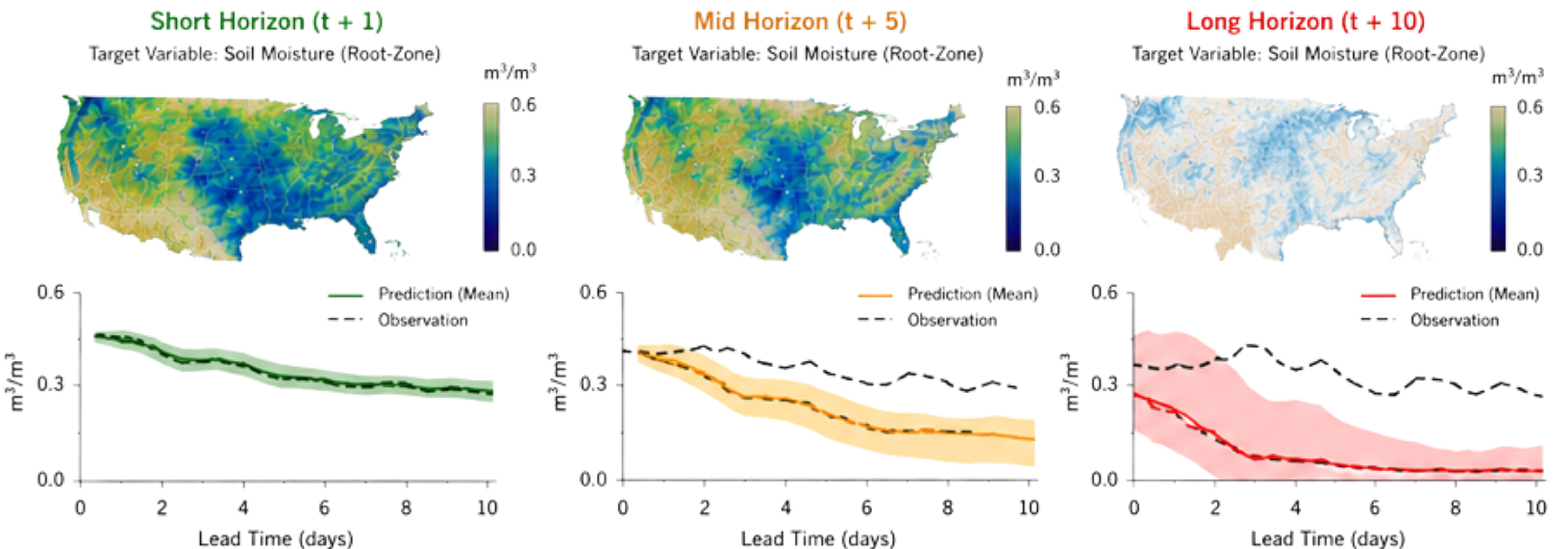}
    \caption{Conceptual illustration of the decline in predictive performance as forecast lead time increases. Spatial prediction maps (top row) and temporal performance curves (bottom row) are shown for short-range ($t+1$), medium-range ($t+5$) and long-range ($t+10$) forecasts. Short-horizon predictions reproduce the target patterns closely, medium-range forecasts drift as errors accumulate and observational constraints weaken, and at long lead times uncertainty dominates.}
    \label{fig:predictive_decline}
\end{figure}

The United Nations Convention to Combat Desertification (UNCCD) defines desertification as land degradation in arid, semi-arid and dry sub-humid regions resulting from climatic variability and human activity \cite{unccd1994}. Unlike most environmental forecasting targets, it has no single directly observable variable behind it. It emerges slowly from the cumulative interaction of several drivers acting together over many years \cite{abdelrahman2023overview,riveramarin2022use}. Left unchecked, these coupled processes erode ecosystem resilience, with consequences for e.g., agricultural productivity, biodiversity, and more. The affected populations become more vulnerable, and the losses extend to carbon storage and socioeconomic stability, touching Zero Hunger (SDG 2), Clean Water and Sanitation (SDG 6), Climate Action (SDG 13) and Life on Land (SDG 15) \cite{un2015agenda,unccd1994,tavares2019desertificacao}.

Northeastern Brazil's semi-arid region is a clear example. Recurrent droughts and vegetation degradation reinforce each other there and drive a steady deterioration of the landscape \cite{souza2023sistemas,dacunto2024rural}. EO has done much to characterize this process as it unfolded: land-cover products and each general information contribute part of the historical picture \cite{riveramarin2022use,dacunto2024rural}. Forecasting where it goes next is considerably harder, since many of the mechanisms driving desertification are observed only indirectly, if at all, or at resolutions that do not match the rest of the record \cite{abdelrahman2023overview,dacunto2024rural}. A forecasting framework meant to inform decisions therefore needs more than accuracy. It needs a fair reference for what is predictable, some notion of physical consistency, uncertainty quantification, and the ability to absorb richer EO data as it becomes available \cite{zhu2026foundations,xiao2025foundation}. These properties matter more as specialized forecasting modules are folded into EO foundation-model ecosystems \cite{zhou2025advances,zhu2026foundations,xiao2025foundation}.

Among deep learning architectures suited to this kind of problem, the Convolutional Long Short-Term Memory network (ConvLSTM) \cite{shi2015convolutional} is a natural starting point. Convolutional operations are embedded inside its recurrent memory cells, so it learns how spatiotemporal patterns evolve across a multivariate EO time series while keeping the spatial structure of the data \cite{shi2015convolutional,shi2017deep}.

We study such a model in the Brazilian state of Cear\'a, a semi-arid testbed with pronounced climatic variability and a well-documented history of desertification \cite{tavares2019desertificacao,souza2023sistemas}. The ConvLSTM is used as an instrument to answer three questions. How much of the year-to-year change in a satellite-derived degradation index can be predicted from the EO record available today? At which horizon, if any, does a learned model add information beyond the persistence of the current state? And what does that limit say about the processes that EO does and does not observe? Our contributions are (i) a model-independent estimate of the predictability of annual degradation change in Cear\'a, (ii) a residual ConvLSTM trained with a curriculum and evaluated against that estimate on a contiguous spatial and temporal holdout, and (iii) the diagnosis of a failure mode in which a network that reconstructs the whole field produces spurious mean reversion when run recursively.

\section{Methodology}

\subsection{Study Area}

The study covers the state of Cear\'a in northeastern Brazil, a semi-arid region with pronounced hydroclimatic variability and extensive land degradation \cite{rolim2026analyzing,silva2017identificacao}. It is one of the country's principal desertification-affected areas. Prolonged droughts coexist with vegetation degradation and land-use pressures such as burning and slash-and-burn agriculture, and these interact across scales to leave a heterogeneous footprint of degradation on the landscape \cite{tavares2019desertificacao,souza2023sistemas,silva2017identificacao,macedo2024everythings}. Cear\'a is a useful testbed because relatively preserved caatinga and severely degraded terrain occur side by side, giving a wide spread of environmental trajectories against which a forecast can be tested \cite{rolim2026analyzing,silva2017identificacao}.

\subsection{Earth Observation Dataset Construction}

All variables were assembled on Google Earth Engine (GEE) \cite{gorelick2017google} through its Python API, aggregated to annual resolution for 2000--2023 and exported onto a common grid of $256 \times 206$ cells in EPSG:4326, covering 41.4\textdegree W--37.3\textdegree W and 7.9\textdegree S--2.8\textdegree S with cells of about 0.02\textdegree{} (roughly 2.2~km). Products with finer native resolution were first brought to 1~km by a maximum-value reduction over the contributing pixels followed by reprojection, and all layers were then resampled to this grid.

Ten variables make up the dataset (Table~\ref{tab:variables}). The forecasting target is the Integrated Desertification Vulnerability (IDV) index, derived from the MapBiomas land use and land cover collection \cite{mapbiomas2024}. Pixels classified as pasture or other non-vegetated area are set to one and all others to zero, and the resulting binary map is smoothed with a Gaussian kernel ($\sigma = 1$ pixel). After aggregation to the analysis grid, the IDV of a cell is a continuous value in $[0,1]$ that expresses the presence and intensity of degradation-prone cover within it. The other nine variables are predictors. Climate and hydrology are described by annual precipitation from CHIRPS \cite{funk2015climate}, mean daytime land surface temperature (LST) from MODIS MOD11A2 \cite{wan2002validation}, and the Standardized Precipitation Index on a 12-month accumulation (SPI-12) \cite{svoboda2012standardized}, obtained from the annual CHIRPS totals by fitting a gamma distribution per pixel. Vegetation is described by NDVI and EVI from MODIS MOD13Q1 \cite{huete2002overview}. Soil moisture is the annual mean volumetric water content of the second ERA5-Land layer (7--28~cm) \cite{munozsabater2021era5}. Because the annual mean was sampled without a fixed projection, this channel was extracted at about 1\textdegree{} and carries only the broad regional pattern of soil moisture. Elevation comes from SRTM \cite{reuter2007evaluation}, and slope and aspect are derived from it on the analysis grid after a $3 \times 3$ smoothing.

Each group of variables has a specific role. Vegetation indices and hydroclimatic variables capture ecosystem response and environmental forcing, both to abrupt events such as El Ni\~no droughts and to slower changes. Terrain variables describe long-term landscape susceptibility, and the IDV carries the accumulated history of degradation. All products are available globally, which keeps the approach reproducible outside Cear\'a.

Cells outside the boundary of Cear\'a were exported with a sentinel value. We define a validity mask as the set of cells with valid IDV in every year, which contains 30{,}257 cells (57.4\% of the grid). Outside the mask the tensor is filled with constant values so that it stays dense, but those cells carry no information. Every loss, metric, statistic and figure reported in this paper is computed over valid cells only. The final dataset is the tensor
\begin{equation}
\mathbf{X} \in \mathbb{R}^{T \times C \times H \times W},
\label{eq:tensor}
\end{equation}
with $T=24$ years, $C=10$ variables, $H=256$ and $W=206$.

\begin{table}[!t]
\renewcommand{\arraystretch}{1.2}
\centering
\footnotesize
\setlength{\tabcolsep}{3pt}
\caption{Variables in the Earth Observation Dataset. All Layers Are Resampled to the Analysis Grid of About 2.2~km. The IDV Is Both the Target and an Input Channel.}
\label{tab:variables}
\begin{tabular}{llll}
\toprule
\textbf{Variable} & \textbf{Source} & \textbf{Native res.} & \textbf{Aggregation} \\
\midrule
IDV (target)  & MapBiomas            & 30\,m            & annual       \\
Precipitation & CHIRPS               & 0.05\textdegree  & annual sum   \\
LST           & MOD11A2              & 1\,km            & annual mean  \\
NDVI          & MOD13Q1              & 250\,m           & annual mean  \\
EVI           & MOD13Q1              & 250\,m           & annual mean  \\
Elevation     & SRTM                 & 30\,m            & static       \\
SPI-12        & CHIRPS               & 0.05\textdegree  & annual index \\
Slope         & SRTM                 & 2.2\,km grid     & static       \\
Aspect        & SRTM                 & 2.2\,km grid     & static       \\
Soil moisture & ERA5-Land, 7--28\,cm & 0.1\textdegree   & annual mean  \\
\bottomrule
\end{tabular}
\end{table}

The historical IDV enters the model as one of the ten input channels, while its future values are the quantity being forecast.

\subsection{Normalization}

Each channel is mapped to $[0,1]$ using the 2nd and 98th percentiles of its valid cells over the training years (2000--2017), and values outside that range are clipped. Restricting the percentiles to valid cells matters here. When the fill values outside the state were included, the lower percentile of NDVI was set by the fill value (0.30, against 0.39 inside the state), precipitation lost the lower 312~mm of its range, and the upper percentile of IDV dropped to 0.967, which clipped 3.5\% of the valid cells in the training years. Those clipped cells are the most degraded ones, so the old statistics distorted exactly the part of the target that matters most.

\subsection{Problem Formulation and Sample Construction}
\label{sec:tensor_window}

Forecasting is formulated as autoregressive sequence prediction. Given the $L = 3$ most recent annual states $\mathbf{X}_{t-2}, \mathbf{X}_{t-1}, \mathbf{X}_{t}$, the model predicts the IDV of year $t+1$. During training, each sample contains an input window of $L=3$ years followed by $K=4$ target years,
\begin{equation}
\mathbf{X}^{(i)} = \mathbf{X}_{t:t+L-1}, \qquad \mathbf{Y}^{(i)} = \mathbf{X}_{t+L:t+L+K-1},
\end{equation}
and the four target years are predicted recursively (Section~\ref{sec:training_strategy}). The $K=4$ training rollout is distinct from the seven-year inference horizon (2024--2030) described in Section~\ref{sec:inference}.

Samples are $64 \times 64$ patches extracted with a stride of 32 cells. The split is both spatial and temporal. Training samples use the years 2000--2017 and patches lying entirely in the northern part of the grid (rows 0--127, north of about 5.35\textdegree S). Validation samples use the years 2015--2023, with targets in 2018--2023, and patches lying entirely in the southern part (rows 192--255, south of about 6.63\textdegree S). The 64 rows between the two blocks are not used, so training and validation patches never touch. This yields 216 training samples and 18 validation samples (six patch positions times three temporal windows). Training samples are augmented with horizontal flips, with aspect transformed to $360\text{\textdegree} - a$ so that it remains physically consistent.

The validation band is small and covers only the far south of the state, including the Cariri region. As shown in Section~\ref{sec:predictability_results}, it also degrades faster than the northern training block. Results on it should be read as a test of transfer to a different and more dynamic region, which is a demanding but narrow test.

\subsection{Residual ConvLSTM Architecture}

The model is a three-layer ConvLSTM with 32, 64 and 32 hidden channels and $3 \times 3$ kernels. Within each layer, the input gate, forget gate, candidate memory and output gate are
\begin{align}
i_t &= \sigma\left(W_{xi} * X_t + W_{hi} * H_{t-1} + b_i\right), \\
f_t &= \sigma\left(W_{xf} * X_t + W_{hf} * H_{t-1} + b_f\right), \\
\tilde{C}_t &= \tanh\left(W_{xc} * X_t + W_{hc} * H_{t-1} + b_c\right), \\
o_t &= \sigma\left(W_{xo} * X_t + W_{ho} * H_{t-1} + b_o\right),
\end{align}
where $*$ denotes convolution and $\sigma$ the logistic function. The cell and hidden states are updated as
\begin{equation}
C_t = f_t \odot C_{t-1} + i_t \odot \tilde{C}_t, \qquad H_t = o_t \odot \tanh(C_t).
\label{eq:cellstate}
\end{equation}
The forget-gate bias is initialized to one. Spatial dropout ($p=0.2$) is applied between layers, and the final hidden state is passed through group normalization.

Because all transformations are convolutional, neighbouring cells exchange information at every step, and spatially coherent structures can emerge without explicit physical parameterization. The memory of the network spans the three-year input window. The recurrent state starts from zero at every forecast step and is not carried from one step to the next. This choice keeps training, validation and inference on the same input regime. It also avoids a subtle problem of carrying the state while the window slides, in which the cell would read the same years more than once and in reversed order.

The output head is residual. A $3\times3$ convolution with 16 channels, a GELU activation and a $1\times1$ convolution map the normalized hidden state to a field $r_t$, and the forecast is
\begin{equation}
\hat{Y}_{t+1} = \mathrm{clip}_{[0,1]}\left(Y_t + \delta_{\max}\tanh(r_t)\right),
\label{eq:residual}
\end{equation}
where $Y_t$ is the most recent IDV in the input window and $\delta_{\max} = 0.6$, close to the 99.99th percentile of the absolute annual change of normalized IDV. The last convolution is initialized at zero, so at the start of training the model is exactly the persistence forecast, and everything it learns is a departure from persistence. The clip uses a straight-through gradient so that cells at the bounds still receive a learning signal. We adopted this form after finding that a model with a sigmoid output, which has to redraw the entire field every year, produced changes unrelated to the observed ones (Section~\ref{sec:change_skill}).

\begin{figure}[!t]
\centering
\includegraphics[width=\columnwidth]{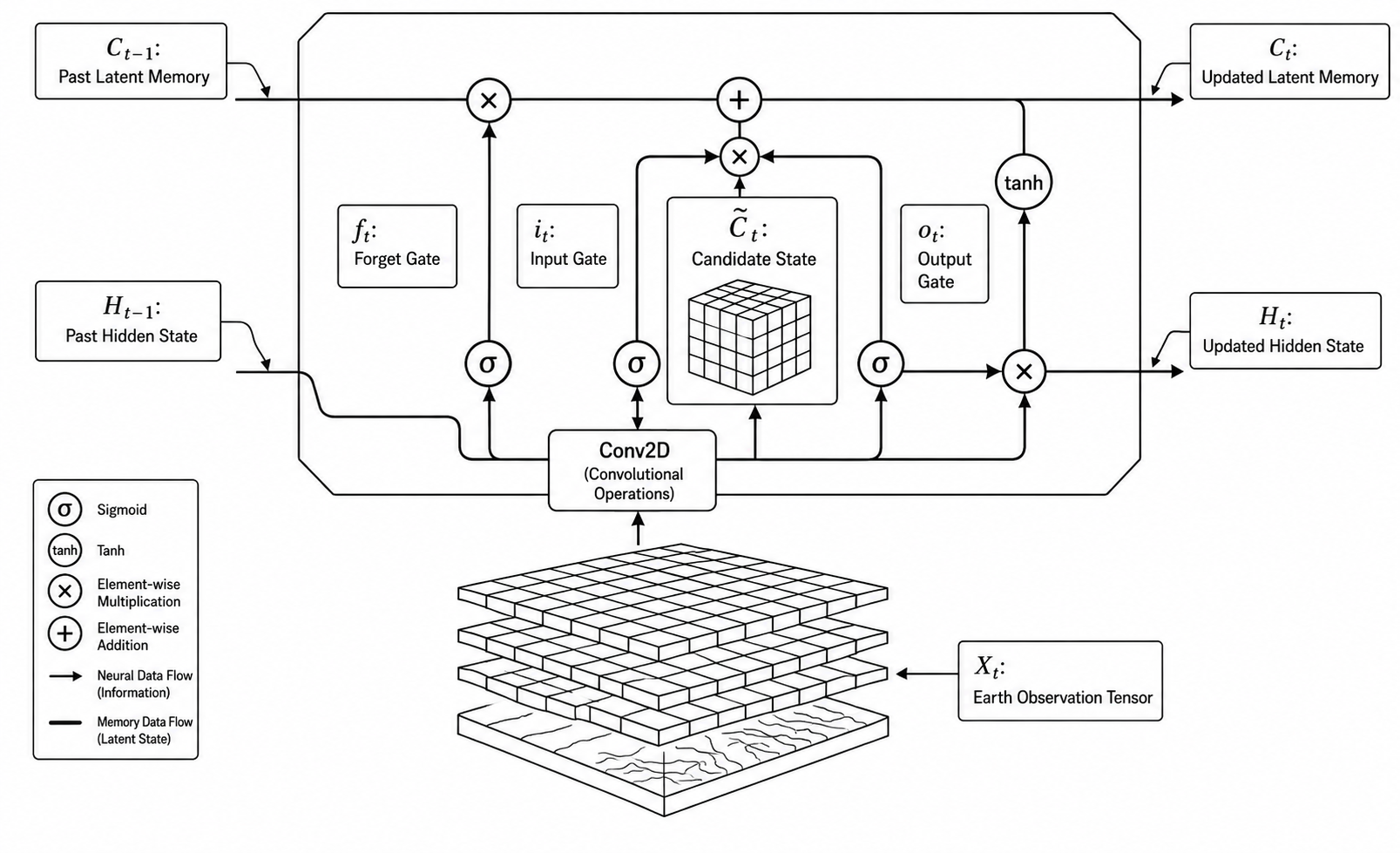}
\caption{The ConvLSTM cell. The input $X_t$ is processed through convolutional gates that update the memory $C_t$ and the hidden state $H_t$. The forget gate $f_t$, input gate $i_t$ and output gate $o_t$ regulate memory retention, the incorporation of new information and the exposure of memory to the output. In this study the memory spans the three-year input window.}
\label{fig:convlstm_cell}
\end{figure}

\subsection{Training Strategy}
\label{sec:training_strategy}

Training uses a masked Huber loss \cite{huber1964robust} with $\delta = 0.1$, computed over valid cells only,
\begin{equation}
\mathcal{L}_\delta(e) =
\begin{cases}
\dfrac{1}{2}e^2, & |e| \le \delta, \\[6pt]
\delta\left(|e| - \dfrac{\delta}{2}\right), & |e| > \delta.
\end{cases}
\end{equation}
The loss is quadratic for ordinary errors and linear for large ones, which limits the influence of isolated classification changes in the MapBiomas record.

The network is trained in two stages, following the idea of curriculum learning \cite{bengio2009curriculum}. In Stage A the model learns the one-step forecast only, from three observed years to the next. It is trained for up to 15 epochs with AdamW \cite{loshchilov2019decoupled} at a learning rate of $10^{-3}$, and the weights with the lowest validation loss are kept. In Stage B these weights are fine-tuned on the full four-step recursive rollout. At step $k$ the model receives the three most recent frames, in which the IDV of earlier steps is either the observed value or the model's own prediction, and the rollout loss is
\begin{equation}
\mathcal{L}_{\text{roll}} = \sum_{k=1}^{K} \gamma^{k-1}\, \mathcal{L}_\delta\!\left(\hat{Y}_{t+k} - Y_{t+k}\right), \qquad \gamma = 0.9.
\end{equation}
The choice between observed and predicted IDV follows scheduled sampling \cite{bengio2015scheduled}. The probability of using the observed value decays linearly from 1.0 to 0.3 over the first 15 epochs and is then kept fixed. Exogenous channels for the target years are the observed ones during training and validation. Stage B uses AdamW with a learning rate of $3\times10^{-4}$, weight decay of $10^{-4}$, gradient clipping at norm 1, batches of four samples, a learning-rate reduction by half after five epochs without improvement, and early stopping with a patience of five epochs counted from the end of the scheduled-sampling ramp. The maximum is 40 epochs. Validation is always fully autoregressive, and the checkpoint with the lowest validation rollout loss is kept.

\subsection{Recursive Inference}
\label{sec:inference}

The 2024--2030 projection starts from the observed window 2021--2023. At each step the model forecasts the IDV of the next year, and a new frame is assembled from that forecast and the exogenous channels. Static channels (elevation, slope, aspect) keep their observed values. The dynamic exogenous channels (precipitation, LST, NDVI, EVI, soil moisture and SPI-12) are set to their per-cell 2000--2023 mean in normalized units. The new frame replaces the oldest one in the window and the process repeats. Cells outside the validity mask are held at their observed value, so the model's output there is never fed back into the window.

The projection is therefore conditional on average climate. It describes how the degradation pattern would evolve if future years resembled the climatological mean, and it should not be read as a climate prediction. Section~\ref{sec:sensitivity} shows that this choice has a small effect on the hindcast.

\subsection{Uncertainty Quantification}
\label{sec:uncertainty_quantification}

We use Monte Carlo (MC) Dropout \cite{gal2016dropout}. The spatial dropout layers of the selected network are kept active at inference, and $D=50$ stochastic recursive rollouts give a mean $\bar{Y}_{\mathrm{MC}}$ and a per-cell standard deviation $\sigma_{\mathrm{MC}}$. MC Dropout approximates the variability of one network around its learned weights, so it should be read as an indicator of relative confidence across space and time, not as a calibrated probability. The 95\% bands are $\bar{Y}_{\mathrm{MC}} \pm 1.96\,\sigma_{\mathrm{MC}}$, clipped to $[0,1]$. The cell-level variability that this spread does not capture is represented by the scenarios of Section~\ref{sec:scenario_methods}.

\subsection{Scenario Maps}
\label{sec:scenario_methods}

The network forecasts the expected IDV of each cell, and the expected value of a process that changes by jumps is smooth. In the observed record about 55\% of the valid cells do not change at all from one year to the next, while the rest move by amounts with a standard deviation of about 0.05. The expected forecast instead moves every cell by a few thousandths per year. Over several recursive steps this keeps the regional mean on track but pulls the cells at both ends of the scale toward the middle, so the maps lose the stable cells at zero and the saturated hotspots that the observed maps keep.

To obtain maps with realistic cell-level behaviour we generate stochastic scenarios around the network forecast. At each step and cell, the change is either zero, with the probability $p_b$ observed for cells of the same IDV level $b$ (eleven classes from 0 to 1), or a jump drawn from the observed non-zero changes of that class. The jumps $J_b$ are shifted by an amount $s$ chosen so that the expected change of every cell, $(1-p_b)\left(\mathbb{E}[J_b]+s\right)$, equals the change $\mu$ forecast by the network, which keeps the scenario mean anchored to the ConvLSTM. Two latent Gaussian fields drive the draws, one deciding whether a cell changes and one setting the size of the jump. Both are smoothed with a Gaussian kernel whose width reproduces the observed spatial correlation of stable cells, and the first follows a first-order autoregression in time whose coefficient reproduces the observed probability that a cell stays stable two years in a row. Each scenario is fed back into the network, so the network sees its own sampled history. The class probabilities and jump distributions are estimated from observed changes only, using the training years for the hindcast and all years for the projection. We run 40 scenarios.

\subsection{Predictability Reference}
\label{sec:predictability_methods}

To judge the network fairly we first measure how predictable the target is, using methods that do not depend on the ConvLSTM.

\emph{Transition statistics.} For all pairs of consecutive years in the training period we compute the annual change $\Delta Y_{t+1} = Y_{t+1} - Y_t$ at every valid cell and summarize it by the level of $Y_t$. The mean of $\Delta Y$ in each class is what an ideal predictor that knows only the current level would forecast.

\emph{Simple baselines.} Starting from the observed 2017 map, we forecast 2018--2023 with (i) persistence, (ii) persistence plus the training-period mean annual change applied $k$ times, (iii) persistence plus the mean change of the current IDV class, iterated year by year, and (iv) a per-cell linear trend fitted to 2000--2017. These baselines are evaluated over all valid cells and over the southern validation band.

\emph{Learned upper reference.} A gradient-boosting regressor \cite{pedregosa2011scikit} (300 iterations, learning rate 0.05, at most 63 leaves, at least 200 samples per leaf) is trained to predict $\Delta Y_{t+1}$ at each cell. Its features are $Y_t$, the two most recent annual changes, the mean of $Y_t$ over $3\times3$, $9\times9$ and $21\times21$ neighbourhoods, the $9\times9$ mean of the most recent change, the values at $t$ and $t+1$ and their difference for each dynamic exogenous channel, and elevation and slope. It is trained on 2002--2017 transitions in the northern training block and tested on 2018--2023 transitions with observed inputs. This reference sees the observed exogenous values of the target year, as the ConvLSTM does during validation, so it is a generous estimate of what the ten channels can explain.

\emph{Change skill.} RMSE, $R^2$ and SSIM between forecast and observed maps are dominated by the spatial pattern, which persistence reproduces almost perfectly. To measure whether a model predicts change, we compare its one-step predicted change $\hat{\Delta}_{t+1} = \hat{Y}_{t+1} - Y_t$ with the observed change through their correlation and their standard deviations. We also scale the predicted change by a factor $\alpha \in [0,1]$ and find the $\alpha$ that minimizes RMSE on training years and rows. A value near one means the predicted changes are informative at their full size, and a value near zero means they only add error.

\subsection{Evaluation Metrics and Consistency Checks}
\label{sec:consistency_methods}

Forecasts are scored over valid cells with RMSE, MAE, the coefficient of determination $R^2$, the Pearson correlation $r$ and the Structural Similarity Index (SSIM) \cite{wang2004image},
\begin{equation}
\mathrm{SSIM}(x,y) = \frac{(2\mu_x\mu_y+C_1)(2\sigma_{xy}+C_2)}{(\mu_x^2+\mu_y^2+C_1)(\sigma_x^2+\sigma_y^2+C_2)},
\end{equation}
computed with an $11\times11$ Gaussian window ($\sigma=1.5$) and averaged over valid cells. RMSE reacts strongly to large local errors, MAE describes the typical error, $R^2$ penalizes bias against the observed variance, $r$ measures linear association only, and SSIM describes local structure. Metrics are computed for each validation batch and forecast step and then averaged. For a persistent field such as the IDV, all five are dominated by the stable spatial pattern, which is why the change-skill diagnostic above is needed.

The recursive projection is also checked for physical plausibility against the historical record, using five criteria computed over valid cells. With $A_t$ the fraction of valid cells whose IDV exceeds the 70th percentile of the historical IDV, the annual change $|A_t - A_{t-1}|$ must stay below 15 percentage points, and the largest predicted one-year increase must not exceed three times the largest historical one (with a floor of 5 points). Hotspots are cells above the historical 80th percentile, and their persistence is measured by the intersection over union between consecutive years,
\begin{equation}
\mathrm{IoU} = \frac{|A \cap B|}{|A \cup B|},
\label{eq:iou}
\end{equation}
which must exceed 0.2. The IDV-weighted centroid of the map,
\begin{equation}
\mathbf{c}_t = \frac{\sum_i Y_{t,i}\,\mathbf{p}_i}{\sum_i Y_{t,i}},
\label{eq:centroid}
\end{equation}
where $\mathbf{p}_i$ is the grid position of cell $i$, must not move more than 30 cells between consecutive years. Finally, the lag-one correlation of the state-wide mean IDV series, historical and projected together, must exceed 0.4. These checks are coarse proxies for plausibility. They can flag a forecast that relocates or inflates degradation implausibly, but passing them does not validate the forecast.

\subsection{Channel Reliance and Zone-Stratified Evaluation}
\label{sec:ablation_methods}

Channel reliance is measured by knockout at inference time. The trained model is evaluated on the validation rollout with one channel, or a group of channels, replaced by the constant 0.5 in all input and appended frames, without retraining. The configurations remove SPI-12, NDVI, EVI, LST, precipitation or soil moisture one at a time, both vegetation indices together, elevation alone, slope and aspect together, the three terrain channels (``Without topography''), or all dynamic exogenous channels, leaving only the IDV and terrain (``Topography only''). The historical IDV is not knocked out, since it is the state being forecast. Because there is no retraining, the increase in RMSE measures how much the trained network relies on a channel, not how much information the channel carries. Each knocked-out run is compared with the full model on the same batches and forecast steps, and the 95\% intervals of the RMSE difference come from 1{,}000 paired bootstrap resamples of these per-batch, per-step differences.

Spatial transfer is examined with five climate zones of Cear\'a defined by bounding boxes: Norte Litoral, Sert\~ao Central, Sul Seco, Maci\c{c}o de Baturit\'e and Cariri. Each patch is assigned to the smallest box that contains its centre, and the trained model is evaluated on the 2015--2023 windows of patches inside and outside each zone. The model is not retrained with zones withheld, so this is a zone-stratified evaluation and not a strict holdout. Because the validation band lies in the south, the in-zone patches of Sul Seco and Cariri come entirely from the validation band, and their gap partly reflects the train--validation split itself.

\section{Results}

\subsection{Observed Dynamics and Predictability}
\label{sec:predictability_results}

Mean IDV over valid cells rose from 0.238 in 2000 to 0.318 in 2017 and 0.350 in 2023, an average of $+0.0047$ per year over the training period and $+0.0055$ per year over 2018--2023. The rise is not uniform. In the northern training block mean IDV went from 0.190 to 0.257 between 2000 and 2023 ($+0.0024$ per year until 2017 and $+0.0045$ afterwards). In the southern validation band it went from 0.257 to 0.401 ($+0.0054$ and $+0.0088$ per year). The validation band therefore degrades faster than the region the model learns from, and faster in the validation years than in the training years.

\begin{table}[!t]
\renewcommand{\arraystretch}{1.2}
\caption{Annual IDV Change by Current IDV Level, Valid Cells. Training Transitions 2000--2017 and Validation Transitions 2017--2023.}
\label{tab:transitions}
\centering
\footnotesize
\setlength{\tabcolsep}{3pt}
\begin{tabular}{lccccc}
\toprule
\textbf{$Y_t$ class} & \textbf{Share} & \textbf{Mean $\Delta Y$} & \textbf{Huber loc.} & \textbf{Mean $\Delta Y$} & \textbf{Unchanged} \\
 & (train) & (train) & (train) & (val) & (train) \\
\midrule
0--0.01     & 0.265 & $+0.0056$ & $+0.0046$ & $+0.0059$ & 0.90 \\
0.01--0.2   & 0.260 & $+0.0072$ & $+0.0055$ & $+0.0084$ & 0.52 \\
0.2--0.5    & 0.228 & $+0.0066$ & $+0.0055$ & $+0.0081$ & 0.33 \\
0.5--0.8    & 0.146 & $+0.0016$ & $+0.0028$ & $+0.0033$ & 0.28 \\
0.8--0.99   & 0.078 & $-0.0040$ & $-0.0018$ & $-0.0031$ & 0.39 \\
0.99--1     & 0.023 & $-0.0042$ & $-0.0036$ & $-0.0042$ & 0.83 \\
\midrule
All         & 1.000 & $+0.0047$ & $+0.0041$ & $+0.0055$ & --- \\
\bottomrule
\end{tabular}
\end{table}

Table~\ref{tab:transitions} shows how this drift is distributed. Cells with low and intermediate IDV gain degradation on average, cells near saturation lose a little, and the median change is zero in every class. Between 28\% and 90\% of the cells in a class do not change at all in a given year (to within $10^{-3}$). The standard deviation of the annual change is 0.053, more than ten times the mean drift. Most of what happens to a single cell from one year to the next is therefore a jump of a few hundredths, up or down, around a slow upward trend. The robust location of the change under the Huber loss used for training ($+0.0041$ per year) is close to the mean ($+0.0047$), so the choice of loss does not by itself remove the drift.

\begin{table}[!t]
\renewcommand{\arraystretch}{1.2}
\caption{RMSE of Simple Baselines Forecasting 2018--2023 from the Observed 2017 Map, by Horizon $k$ in Years.}
\label{tab:baselines}
\centering
\footnotesize
\setlength{\tabcolsep}{2.5pt}
\begin{tabular}{lcccccc}
\toprule
\textbf{Baseline} & $k=1$ & $k=2$ & $k=3$ & $k=4$ & $k=5$ & $k=6$ \\
\midrule
\multicolumn{7}{l}{\emph{All valid cells}} \\
Persistence              & 0.0513 & 0.0759 & 0.0954 & 0.1104 & 0.1282 & 0.1453 \\
+ global drift           & 0.0514 & 0.0760 & 0.0952 & 0.1094 & 0.1259 & 0.1413 \\
+ level-dependent drift  & 0.0514 & 0.0757 & 0.0946 & 0.1083 & 0.1242 & 0.1389 \\
Per-cell linear trend    & 0.0536 & 0.0813 & 0.1042 & 0.1222 & 0.1412 & 0.1589 \\
\midrule
\multicolumn{7}{l}{\emph{Southern validation band}} \\
Persistence              & 0.0514 & 0.0746 & 0.0967 & 0.1144 & 0.1348 & 0.1578 \\
+ global drift           & 0.0510 & 0.0738 & 0.0953 & 0.1118 & 0.1304 & 0.1506 \\
+ level-dependent drift  & 0.0507 & 0.0729 & 0.0937 & 0.1095 & 0.1273 & 0.1465 \\
Per-cell linear trend    & 0.0531 & 0.0790 & 0.1038 & 0.1239 & 0.1457 & 0.1683 \\
\bottomrule
\end{tabular}
\end{table}

Table~\ref{tab:baselines} turns these statistics into forecasts. Over the whole state, adding the drift to persistence makes no difference one year ahead and starts to help at three years. At six years the level-dependent drift lowers RMSE by 4.4\% relative to persistence and reduces the bias in mean IDV from $-0.033$ to $-0.005$. In the faster-degrading southern band the drift helps earlier, by 1.4\% at one year and 7.2\% at six years, although even there it recovers only half of the six-year bias ($-0.053$ for persistence, $-0.027$ with the drift), since that band degrades faster than the training-period rate. Extrapolating each cell's own linear trend is worse than persistence at every horizon, because cell-level trends over 18 years mostly fit noise.

The gradient-boosting reference sets an upper bound on what the ten channels can explain at the cell level. Trained on the northern block and tested on 2018--2023, it explains $R^2 = 0.047$ of the variance of next-year change over the state and $R^2 = 0.012$ in the southern band. This corresponds to an RMSE 2.9\% and 1.9\% below persistence. The changes it predicts have a standard deviation of 0.014, against 0.052 for the observed changes, and its mean predicted change ($+0.0004$ per year) misses most of the observed drift ($+0.0055$). About 95\% of the year-to-year variation of the IDV at the 2.2~km scale is therefore not explained by the information available in these EO products, and the slow regional drift is the most forecastable part of the signal.

\subsection{Forecast Skill Against Persistence}
\label{sec:skill}

\begin{table}[!t]
\renewcommand{\arraystretch}{1.2}
\caption{Single-Step ($k=1$) Performance on the Validation Band, Averaged over Validation Patches}
\label{tab:performance}
\centering
\footnotesize
\setlength{\tabcolsep}{3pt}
\begin{tabular}{lccccc}
\toprule
\textbf{Model} & \textbf{RMSE} & \textbf{MAE} & \textbf{$R^2$} & \textbf{Pearson} & \textbf{SSIM} \\
\midrule
Persistence & 0.0508 & \textbf{0.0250} & 0.975 & 0.988 & \textbf{0.974} \\
CNN         & 0.0546 & 0.0375 & 0.971 & 0.986 & 0.958 \\
LSTM        & 0.0626 & 0.0437 & 0.962 & 0.981 & 0.948 \\
Residual ConvLSTM & \textbf{0.0502} & 0.0274 & \textbf{0.975} & \textbf{0.988} & 0.971 \\
\bottomrule
\end{tabular}
\end{table}

\begin{table}[!t]
\renewcommand{\arraystretch}{1.2}
\caption{Multi-Step RMSE on the Southern Validation Band by Forecast Horizon $k$ (Full-Scene Rollouts Started in 2018, 2019 and 2020). Bottom Rows: Relative RMSE Change of the ConvLSTM, with 95\% Spatial Block-Bootstrap Intervals}
\label{tab:multistep}
\centering
\footnotesize
\setlength{\tabcolsep}{3pt}
\resizebox{\columnwidth}{!}{%
\begin{tabular}{lcccc}
\toprule
\textbf{Model} & $k=1$ & $k=2$ & $k=3$ & $k=4$ \\
\midrule
Persistence         & 0.0514 & 0.0769 & 0.0999 & 0.1228 \\
Persistence + drift & 0.0512 & 0.0764 & 0.0988 & 0.1206 \\
CNN                 & 0.0546 & 0.0802 & 0.1037 & 0.1259 \\
LSTM                & 0.0651 & 0.0838 & 0.1054 & 0.1273 \\
Residual ConvLSTM   & \textbf{0.0509} & \textbf{0.0755} & \textbf{0.0969} & \textbf{0.1173} \\
\midrule
\multicolumn{5}{l}{\emph{ConvLSTM relative to}} \\
Persistence (\%)    & $-1.0$ & $-1.9$ & $-3.0$ & $-4.4$ \\
                    & {\scriptsize[$-1.3,-0.6$]} & {\scriptsize[$-2.5,-1.2$]} & {\scriptsize[$-4.0,-2.1$]} & {\scriptsize[$-5.7,-3.1$]} \\
Pers.\ + drift (\%) & $-0.6$ & $-1.2$ & $-1.9$ & $-2.7$ \\
                    & {\scriptsize[$-0.8,-0.3$]} & {\scriptsize[$-1.6,-0.7$]} & {\scriptsize[$-2.5,-1.3$]} & {\scriptsize[$-3.6,-2.0$]} \\
\bottomrule
\end{tabular}}
\end{table}

The CNN and LSTM baselines take the same three-year input. The CNN stacks the input years as channels and applies two convolutional layers, and the LSTM processes each cell's time series independently. Both use a sigmoid output and are trained for 20 epochs on the one-step task, then run recursively on the validation rollout in the same way as the ConvLSTM. We also include persistence plus a constant drift, equal to the mean annual change of the IDV over the training years and rows (0.0024 per year in normalized units), since Section~\ref{sec:predictability_results} shows that the drift is the most forecastable part of the signal.

Significance is assessed on full-scene rollouts, where each model is run over the whole grid with the outside of the state held fixed. The RMSE difference between the ConvLSTM and each reference is resampled with a paired spatial block bootstrap: the valid cells are grouped into 16~$\times$~16 blocks (24 blocks in the southern band, 147 over the state), and the blocks are drawn with replacement 5000 times while keeping the cell errors of both models together. Blocks of 8 and 32 cells were used to check that the intervals do not depend on the block size. The bootstrap accounts for spatial correlation within a block but not for the fact that all cells share the same years, so we also report the result for each target year separately.

\begin{figure}[!t]
\centering
\includegraphics[width=\columnwidth]{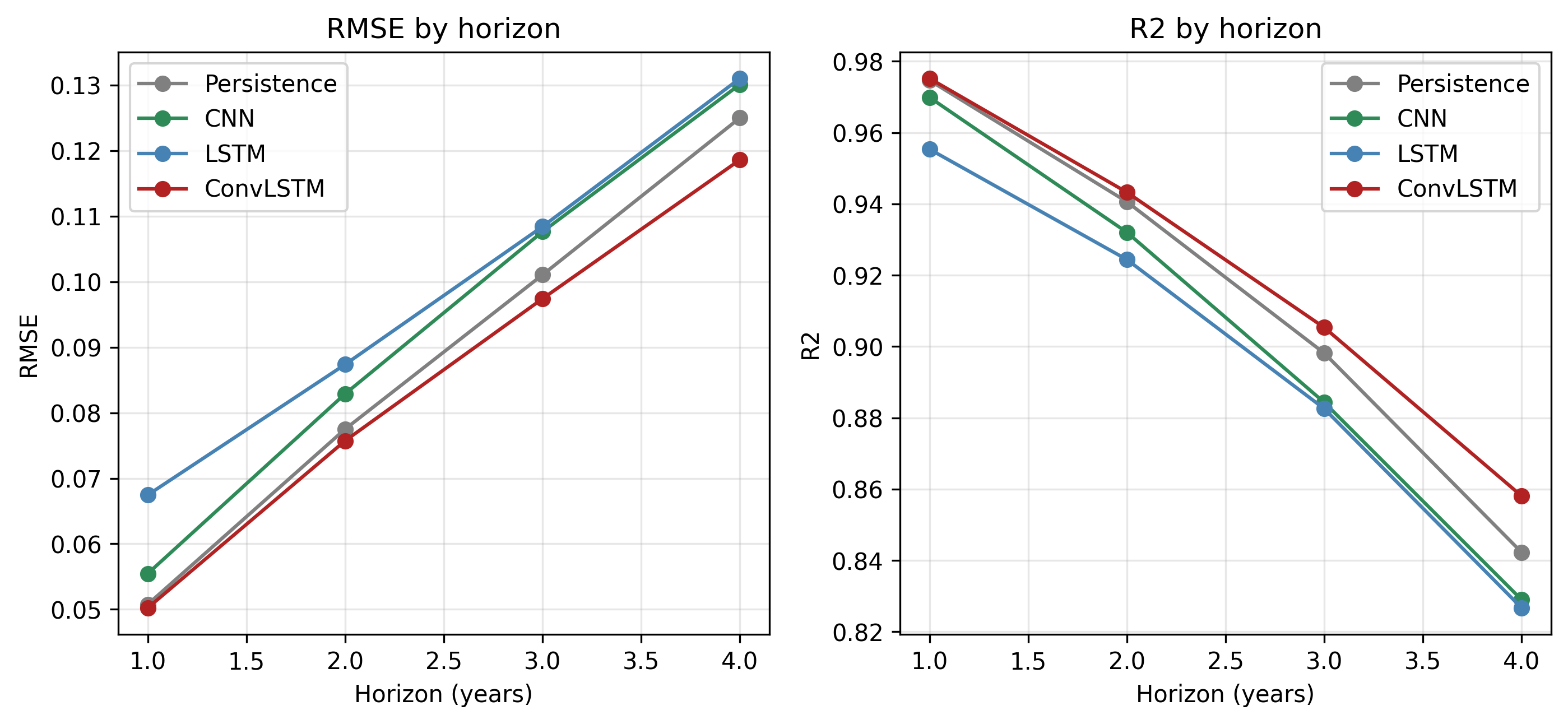}
\caption{RMSE and $R^2$ by forecast horizon for persistence, CNN, LSTM and the residual ConvLSTM on the validation band, averaged over validation patches. Values differ slightly from the full-scene rollouts of Table~\ref{tab:multistep}.}
\label{fig:horizon_benchmark}
\end{figure}

Table~\ref{tab:performance} reports one-step performance on the validation band and Table~\ref{tab:multistep} extends it to four years (see also Fig.~\ref{fig:horizon_benchmark}). The residual ConvLSTM has the lowest RMSE at every horizon. One year ahead it improves on persistence by 1.0\% for the rollouts started in 2018--2020 and by 1.6\% (95\% interval 1.1--2.0\%) when all six target years from 2018 to 2023 are pooled. The gain grows with lead time, to 3.0\% at $k=3$ and 4.4\% at $k=4$, which is the pattern expected if most of what the network adds is the regional drift. It is not only the drift, however. The network also beats persistence with a constant drift at every horizon, by 0.6\% at $k=1$ and 2.7\% at $k=4$, and none of the 5000 bootstrap replicates reversed the sign of either comparison at any horizon ($p < 0.001$). With blocks of 8 and 32 cells, the one-year interval against persistence becomes 1.3--1.9\% and 0.9--2.2\%, so the conclusion does not depend on the block size. Over the whole state, where the drift is slower, the improvement on persistence is smaller: 1.2\% one year ahead (all target years) and 3.4\% at $k=4$. The one-year gain is positive in each of the six target years, both in the southern band (from 0.8\% in 2019 to 2.6\% in 2023) and over the state (from 0.7\% to 1.7\%), and it is largest in 2022 and 2023, the years in which persistence makes its largest errors. These one-year gains should be read against the ceiling of Section~\ref{sec:predictability_results}. The gradient-boosting reference improved on persistence by 2.9\% over the state and 1.9\% in the southern band, so the ConvLSTM recovers about 40\% of the available one-year skill at the state scale and most of it in the south.

The CNN and LSTM baselines are worse than persistence at every horizon. One year ahead their RMSE is about 6\% (CNN) and 27\% (LSTM) above that of persistence, and the gap narrows to 2.5\% (CNN) and 3.7\% (LSTM) at $k=4$ as the persistence error grows. Both redraw the whole field through a sigmoid output, and an output error of a few hundredths of IDV is already as large as the annual change they are meant to forecast. All $R^2$ and SSIM values are high, including those of persistence (0.975 and 0.974 at $k=1$), because they mostly measure how well the stable spatial pattern is reproduced. Differences in skill are better judged from RMSE relative to persistence and from the change diagnostics in the next subsection. Table~\ref{tab:performance} also shows that the ConvLSTM has a higher MAE than persistence (0.0274 against 0.0250) and a slightly lower SSIM. This is expected for a model that adds a small drift everywhere. The median annual change of a cell is zero, so persistence is exactly right for many cells, and a small predicted change on those cells adds to the absolute error while the larger errors elsewhere, which dominate the RMSE, are reduced. The RMSE values in Table~\ref{tab:performance} are averages over validation patches and differ in the fourth decimal from the pooled full-scene values of Table~\ref{tab:multistep}.

\subsection{Does the Network Predict Change?}
\label{sec:change_skill}

An earlier version of the model used a sigmoid output layer, so that it had to reproduce the whole IDV field every year, and carried its recurrent state from one forecast step to the next. Over the four-year validation rollout it reached $R^2$ and SSIM values of about 0.88, which look reasonable for a multi-year forecast. Its one-step predicted changes, however, were unrelated to the observed ones. The correlation between predicted and observed annual change was $-0.03$ over 2003--2017 and $-0.06$ over 2018--2022. When the predicted change was scaled by a factor $\alpha$, the RMSE on training years and rows was lowest at $\alpha = 0$, which means the best use of its predictions was to discard the change and keep persistence. Its errors were small relative to the field but about as large as the real annual changes (standard deviation of about 0.04 against 0.05), and they were not random in their effect. Run recursively, they pulled intermediate and high IDV values downward. In a 2018--2023 hindcast with observed climate, the model's mean IDV fell by 0.006 while the observed mean rose by 0.033. Its 2024--2030 projection dropped by 0.014 in the first year and declined again after 2026, a trajectory that has no support in the record.

The residual formulation of Eq.~(\ref{eq:residual}) was introduced to remove this failure mode, since any change the network outputs must now be learned explicitly. For the residual ConvLSTM, the correlation between predicted and observed one-step change is about 0.13 on training years and rows, and it stays at that level on the validation years, both over the state and in the southern band. For comparison, the $R^2$ of the gradient-boosting reference corresponds to a correlation of about 0.22 over the state and 0.11 in the southern band. The predicted changes are small, with a standard deviation of 0.004--0.005 against 0.052 observed and 0.014 for the gradient-boosting reference, so the network commits only to the part of the change it can support. The RMSE-minimizing scale factor on training data is $\alpha = 1$, the largest value tested, which means its predicted changes are worth keeping in full and are not inflated. Over the validation rollout, the model reproduces 66\%, 69\%, 62\% and 53\% of the observed drift in mean IDV at $k=1$ to 4, always with the correct sign. The CNN, in contrast, captures the drift at $k=1$ and loses it with each recursive step, reaching $-22\%$ at $k=4$. The network has therefore learned a small but real part of the predictable change, consistent in training and validation years, and it under-predicts the drift in the south, as the drift baselines do, because that region degraded faster than the training region.

\subsection{Sensitivity Checks}
\label{sec:sensitivity}

Three design choices were checked with the earlier model, since they concern the evaluation protocol more than the network. First, replacing the observed 2018--2023 exogenous channels with their climatology in the hindcast changed the bias of mean IDV in 2023 only from $-0.038$ to $-0.031$. Climatological forcing is therefore not what limits the forecast, and its use in the 2024--2030 projection costs little. Second, the network's output outside the state boundary is unconstrained by the loss. When it was fed back during recursive inference, the IDV outside the state rose from about 0.03 to 0.18 by 2030 and the error propagated inward from the boundary. Holding those cells at their observed value removed the artefact and raised the projected 2030 mean over valid cells from 0.327 to 0.337. Third, restarting the recurrent state at each step, with the same weights, changed the projected trajectory from a decline after 2026 to a slow rise (0.336 in 2024 to 0.349 in 2030). The current model is trained and evaluated with both corrections from the start, so these numbers motivate the protocol and are not results of the final model.

\subsection{Channel Reliance}
\label{sec:ablation}

\begin{table}[!t]
\renewcommand{\arraystretch}{1.2}
\caption{Channel Knockout on the Validation Rollout ($k=1$--$4$), No Retraining. Change in RMSE Relative to the Full Model (0.0855), with Paired 95\% Bootstrap Intervals}
\label{tab:ablation}
\centering
\footnotesize
\setlength{\tabcolsep}{3pt}
\begin{tabular}{lc}
\toprule
\textbf{Configuration} & \textbf{$\Delta$RMSE [95\% CI]} \\
\midrule
Without SPI-12 & $+0.00010$ [$-0.00001$, $+0.00026$] \\
Without NDVI & $-0.00007$ [$-0.00017$, $+0.00001$] \\
Without EVI & $-0.00045$ [$-0.00085$, $-0.00013$] \\
Without LST & $+0.00052$ [$+0.00021$, $+0.00091$] \\
Without precip. & $+0.00007$ [$0.00000$, $+0.00015$] \\
Without vegetation & $-0.00045$ [$-0.00096$, $-0.00004$] \\
Without soil moisture & $-0.00006$ [$-0.00011$, $-0.00002$] \\
Without elevation & $-0.00028$ [$-0.00049$, $-0.00013$] \\
Without slope, aspect & $-0.00005$ [$-0.00012$, $+0.00003$] \\
Without topography & $-0.00023$ [$-0.00038$, $-0.00010$] \\
Topography only & $-0.00002$ [$-0.00012$, $+0.00008$] \\
\bottomrule
\end{tabular}
\end{table}

\begin{figure}[!t]
\centering
\includegraphics[width=\columnwidth]{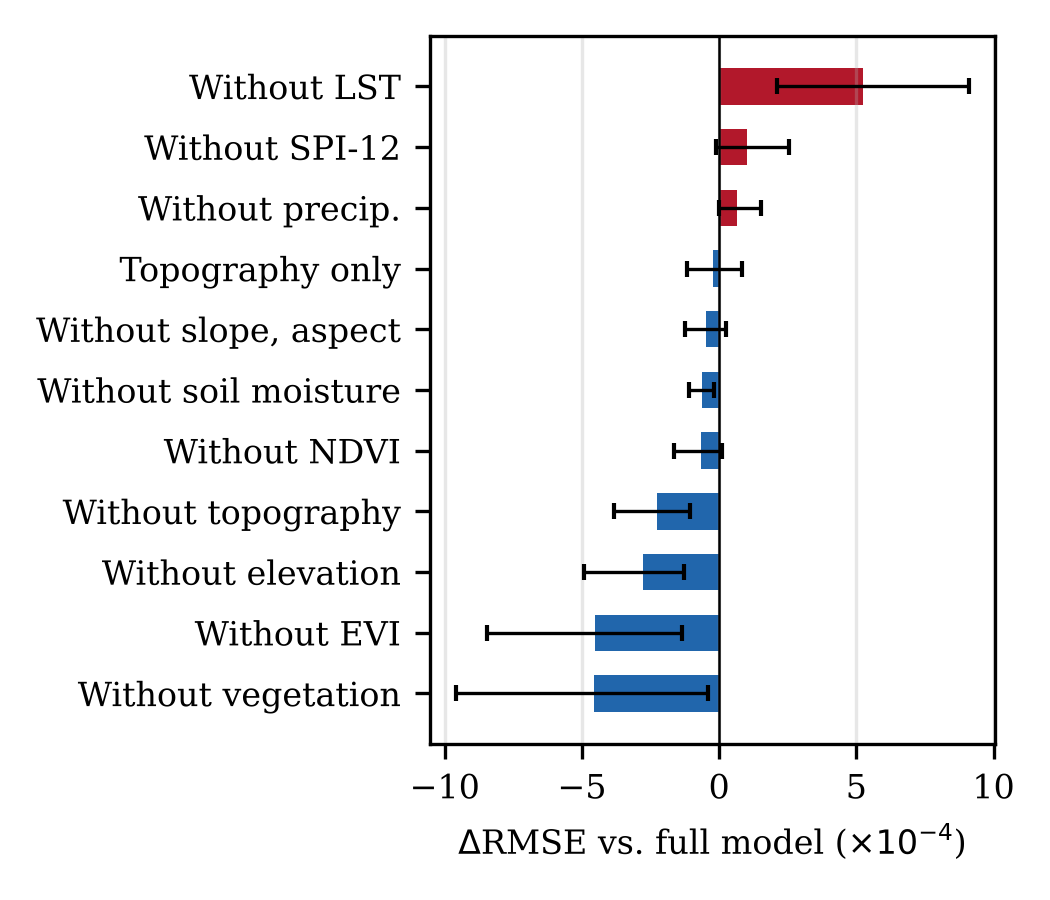}
\caption{Change in validation RMSE when channels are replaced by a neutral constant, with paired 95\% bootstrap intervals. Positive values mean the network relies on the channel. Bars in blue indicate channels whose removal lowers the RMSE.}
\label{fig:ablation}
\end{figure}

Table~\ref{tab:ablation} and Fig.~\ref{fig:ablation} show how much the trained network relies on each channel. Every knockout changes the four-step RMSE by less than 0.001, below 1\% of its value, but the paired intervals separate several of these effects from zero. Replacing LST raises the RMSE by about 0.0005, with an interval that excludes zero, and it is the only channel whose removal clearly hurts. SPI-12 and precipitation raise it by about 0.0001, with intervals that touch zero. Removing EVI, or all the vegetation indices together, lowers the RMSE by about 0.0005, and removing elevation lowers it by about 0.0003, both with intervals that exclude zero. Removing soil moisture also lowers it, but only by 0.00006. Slope and aspect have no measurable effect, and keeping only the topographic channels, which removes all the dynamic inputs at once, leaves the RMSE practically unchanged. In other words, EVI and elevation make the forecast slightly worse in the southern band. Both point to the same problem. The network was trained on the northern block and the relations it learned there between these channels and IDV change do not transfer to the south, which is higher on average, with the Araripe plateau, and degraded faster over the validation period. Elevation in particular can act as a proxy for location, and the expected change of Fig.~\ref{fig:change_map} is largest along the escarpments of the Ibiapaba and Araripe plateaus, where elevation changes most sharply. A model retrained without these channels could test this directly. With a residual head that starts from persistence, effects of this size are what one expects: the predicted changes have a standard deviation of about 0.005, so no single channel can move the RMSE by much. These numbers describe the trained network. A channel the network ignores may still carry information that a different model could use, and a channel it relies on may be standing in for another correlated one.

\subsection{Zone-Stratified Evaluation}
\label{sec:spatial_generalization}

\begin{figure}[!t]
\centering
\includegraphics[width=\columnwidth]{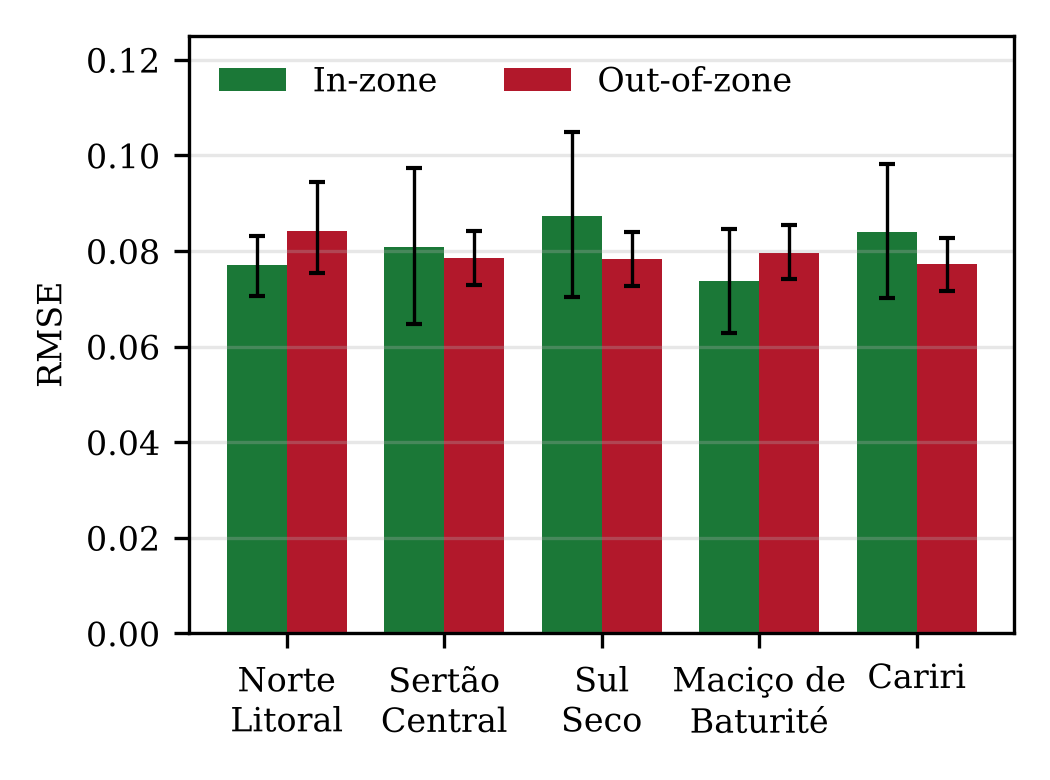}
\caption{RMSE of the residual ConvLSTM on patches inside and outside each climate zone, target years 2018--2023. Error bars are 95\% bootstrap intervals. The model is not retrained per zone.}
\label{fig:spatial_generalization}
\end{figure}

Fig.~\ref{fig:spatial_generalization} compares in-zone and out-of-zone RMSE for the five climate zones. In-zone RMSE ranges from 0.074 in the Maci\c{c}o de Baturit\'e to 0.087 in Sul Seco. It is lower than the out-of-zone RMSE in Norte Litoral (0.077 against 0.084) and the Maci\c{c}o de Baturit\'e (0.074 against 0.080), close to it in the Sert\~ao Central (0.081 against 0.079), and higher in the two southern zones, Sul Seco (0.087 against 0.078) and Cariri (0.084 against 0.077), where the IDV also changed fastest over the validation years. The bootstrap intervals of the in-zone RMSE are wide, from about $\pm 0.006$ in Norte Litoral to $\pm 0.017$ in Sul Seco, and the in-zone and out-of-zone intervals overlap in every zone, so none of these gaps is clearly separated from zero. Patches in Sul Seco and Cariri come entirely from the validation band, and the other zones lie in the training block, so the larger error in the south combines regional differences in the rate of change with the effect of moving from training to validation patches. These results describe how error varies across regions. They are not evidence of generalization to withheld zones.

\subsection{Projection for 2024--2030}
\label{sec:dynamics}

\begin{figure}[!t]
\centering
\includegraphics[width=\columnwidth]{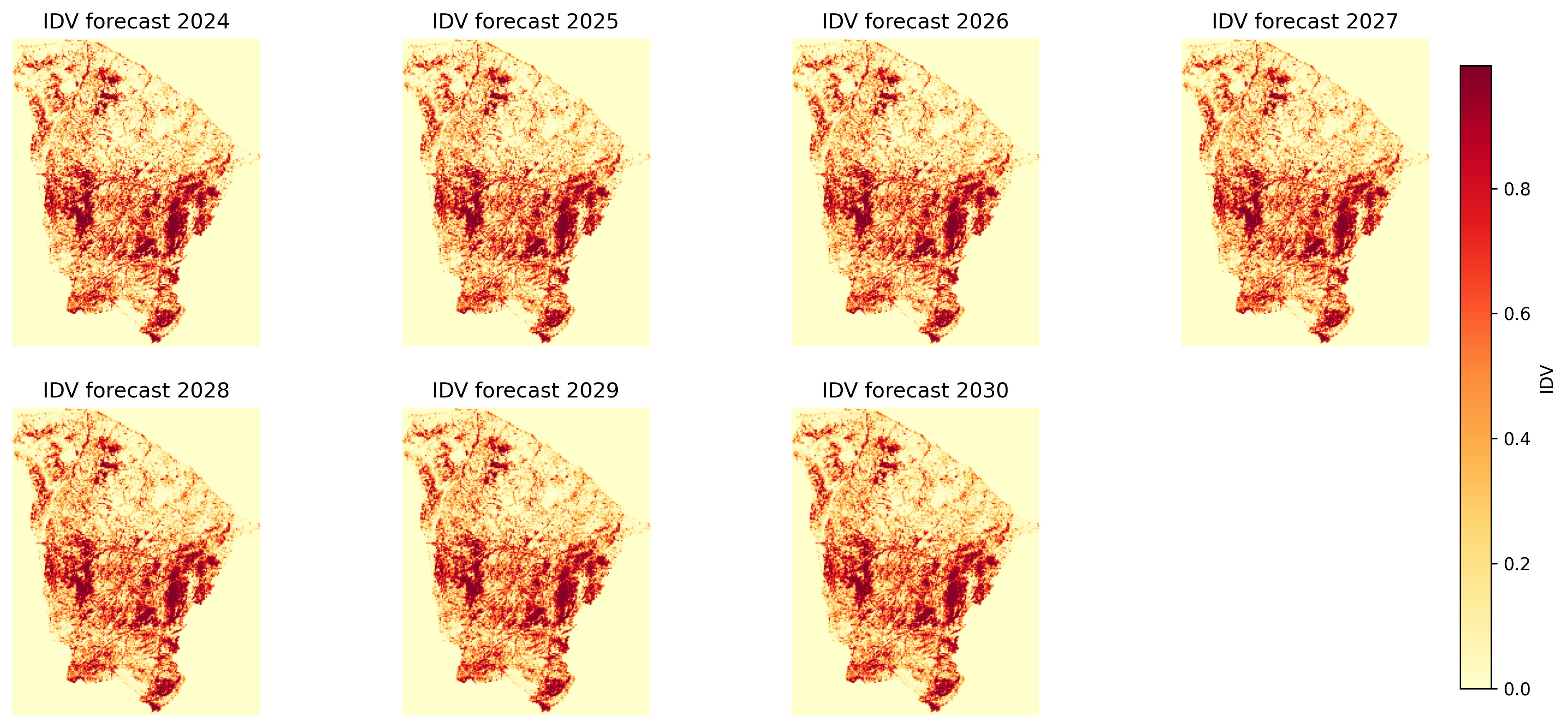}
\includegraphics[width=\columnwidth]{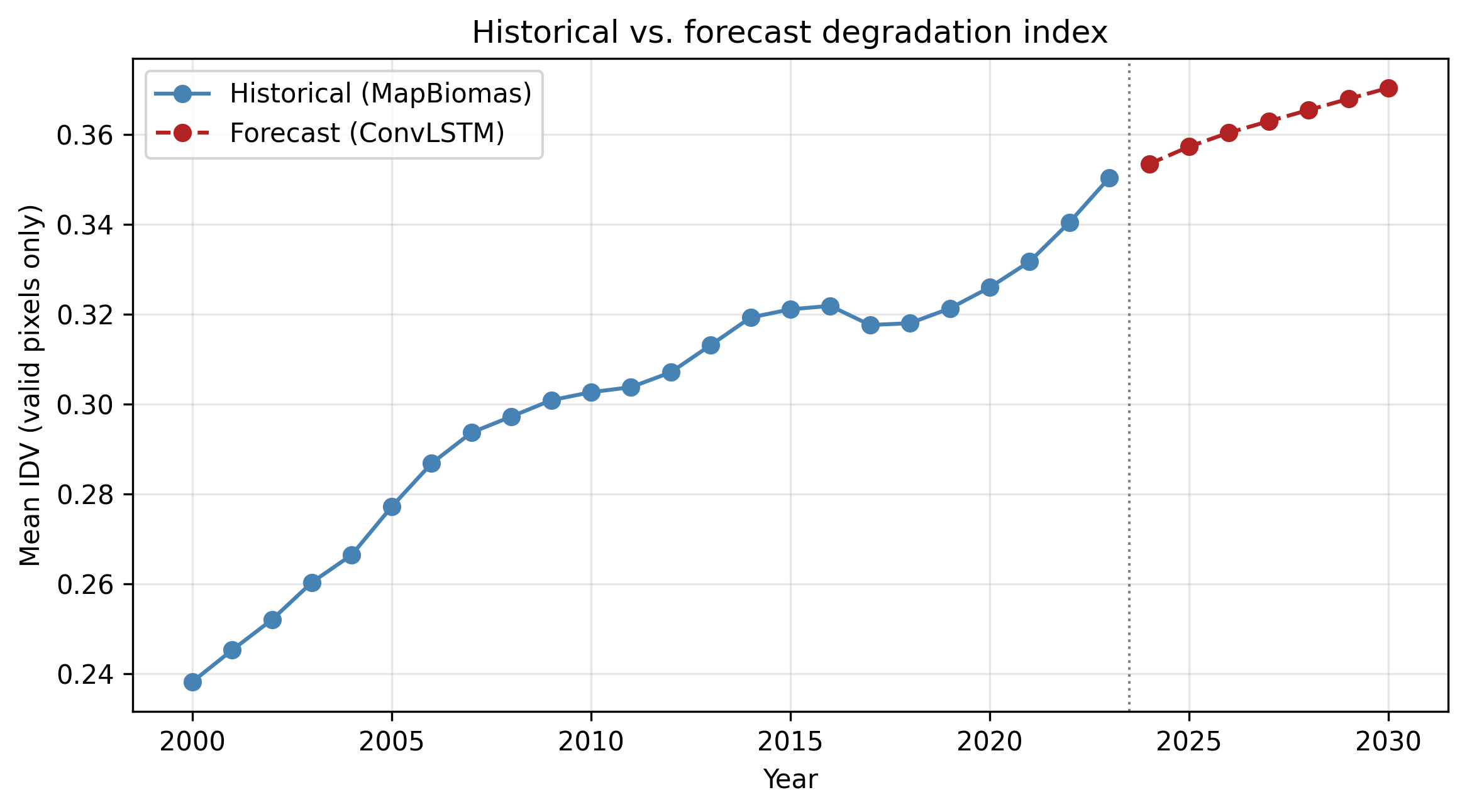}
\caption{Top: recursive projection of the IDV for 2024--2030 under climatological forcing, cells outside Cear\'a held fixed. Bottom: mean IDV over valid cells, observed (2000--2023) and projected (2024--2030).}
\label{fig:forecast_maps}
\end{figure}

Fig.~\ref{fig:forecast_maps} shows the projected maps and the state-wide mean. Under climatological forcing, the mean IDV over valid cells goes from 0.350 in 2023 to 0.370 in 2030, an average change of $+0.0028$ per year against $+0.0055$ per year over 2018--2023 and $+0.0049$ per year over 2000--2023. The projection continues the observed rise but at about half the recent rate, which matches the share of the drift the model reproduced in the hindcast (53--69\%). Taken together with the drift baselines, this suggests that the 2030 value is more likely an underestimate than an overestimate, provided the post-2017 acceleration continues. The spatial pattern stays anchored to the observed one, with the most degraded areas of 2023 remaining the main centres of degradation. This persistence agrees with observations across drylands, where degraded landscapes tend to remain the most susceptible to further deterioration \cite{ibrahim2015land,li2021drivers}. Given Section~\ref{sec:predictability_results}, the projected maps should be read mainly as the 2023 pattern carried forward with a regional trend. They do not identify which individual cells will degrade next.

\begin{figure*}[!t]
\centering
\includegraphics[width=\textwidth]{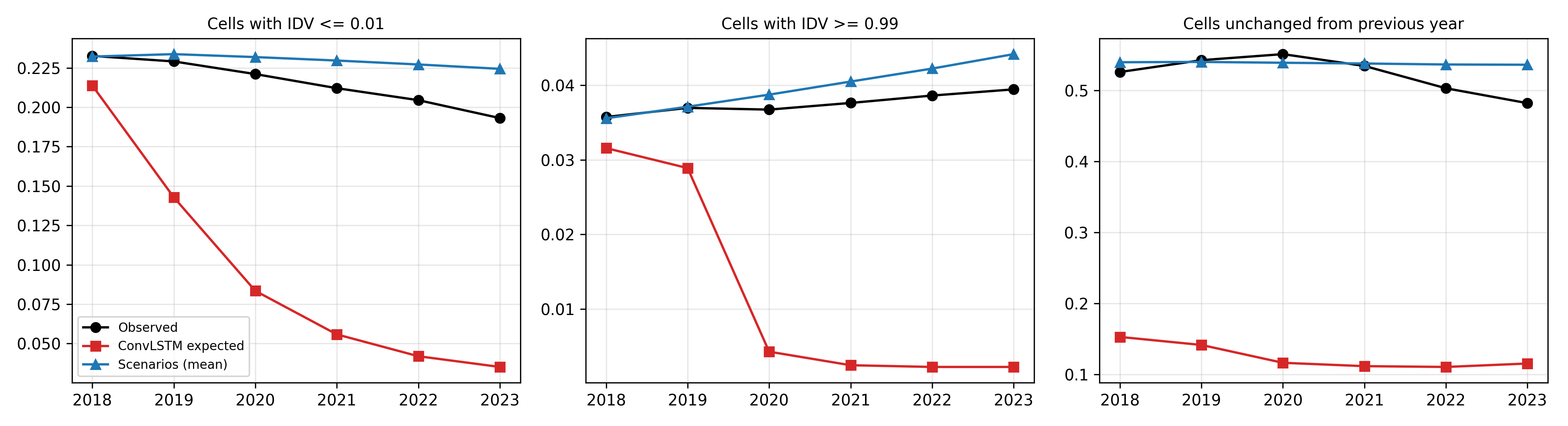}
\caption{Hindcast 2018--2023. Share of valid cells with IDV $\le 0.01$ (left), with IDV $\ge 0.99$ (centre) and unchanged from the previous year (right), for the observed maps, the expected ConvLSTM forecast and the mean of 40 scenarios.}
\label{fig:scenario_hindcast}
\end{figure*}

\begin{figure}[!t]
\centering
\includegraphics[width=\columnwidth]{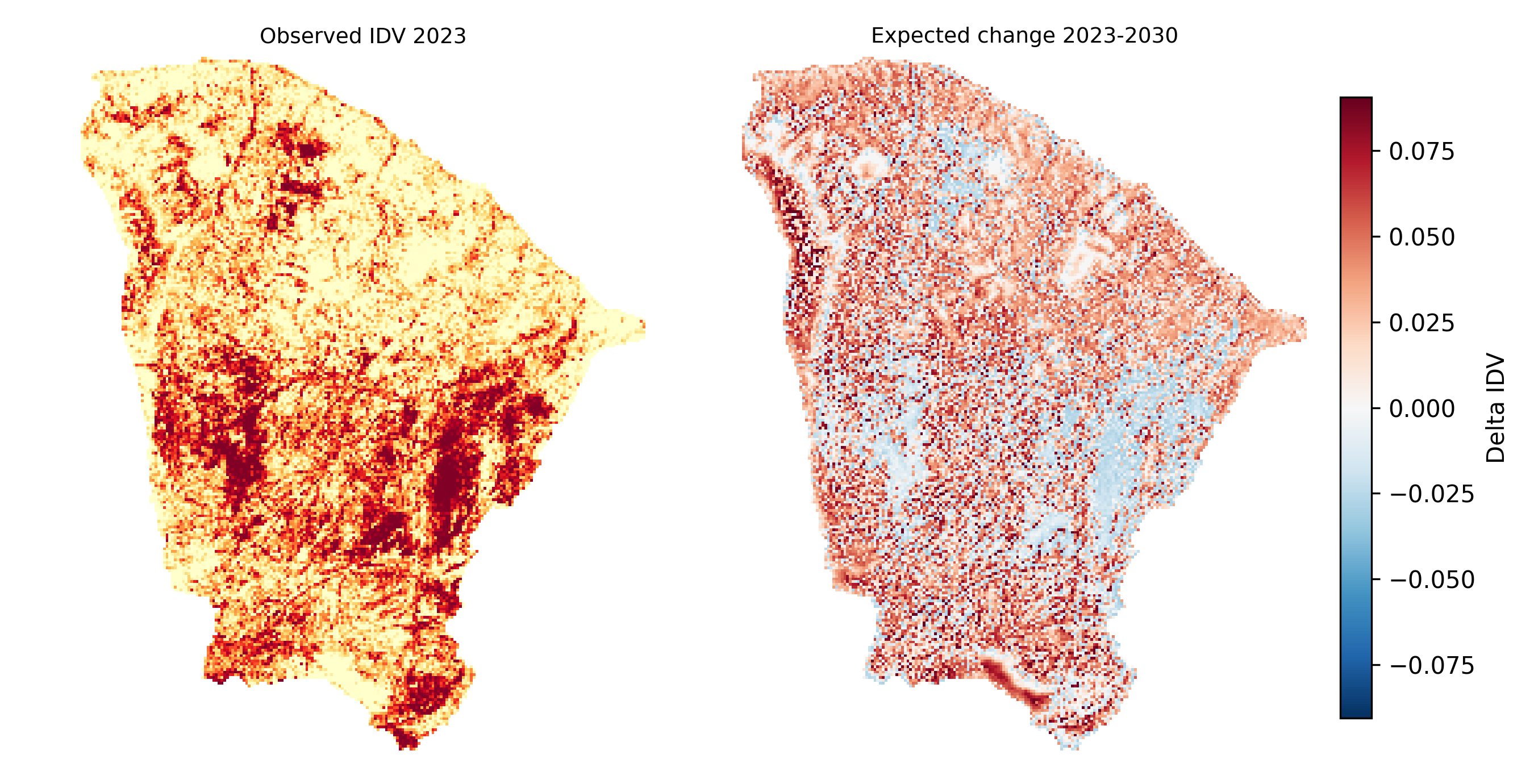}
\caption{Observed IDV in 2023 (left) and expected change from 2023 to 2030 in the ConvLSTM projection (right).}
\label{fig:change_map}
\end{figure}

Because the expected forecast changes every cell slightly, its year-to-year maps look almost identical, and Fig.~\ref{fig:change_map} shows the projected change directly. The change is largest and positive in cells with low IDV and slightly negative in the most degraded cells. This reflects the average behaviour of the record, where cells at low levels gain on average and saturated cells lose, but in the observed maps that average comes from a minority of cells that jump while most stay put. Applied as an expected value to every cell, it erodes both ends of the distribution: in the expected projection the share of valid cells with IDV $\le 0.01$ falls from 0.193 in 2023 to 0.021 in 2030 and the share with IDV $\ge 0.99$ from 0.039 to 0.005. The same happens in the hindcast, where the expected forecast loses most of both groups within three years while the observed shares change by only a few points (Fig.~\ref{fig:scenario_hindcast}).

The scenarios of Section~\ref{sec:scenario_methods} avoid this. In the 2018--2023 hindcast, with the jump model estimated on training years only, 54\% of the cells stay unchanged each year in the scenarios, against 48--55\% observed and 11--15\% for the expected forecast. The share of cells at zero stays between 0.22 and 0.23, against 0.23 to 0.19 observed, and the share of saturated cells rises from 0.036 to 0.044, against 0.036 to 0.039 observed. The scenarios therefore keep the shape of the observed distribution, but they lose zeros a little too slowly and gain saturated cells a little too fast, which matches the faster degradation of the validation period at low levels than in the training years. Their 90\% intervals contain 87--88\% of the observed values, so they are slightly too narrow. The scenarios are also the better probabilistic forecast: their continuous ranked probability score is 20\% lower than the mean absolute error of the expected forecast one year ahead (0.021 against 0.027) and 26\% lower after six years (0.069 against 0.093). Their mean is almost as accurate as the expected forecast (RMSE 0.140 against 0.139 after six years), as it should be, since both are anchored to the same network output.

\begin{figure*}[!t]
\centering
\includegraphics[width=\textwidth]{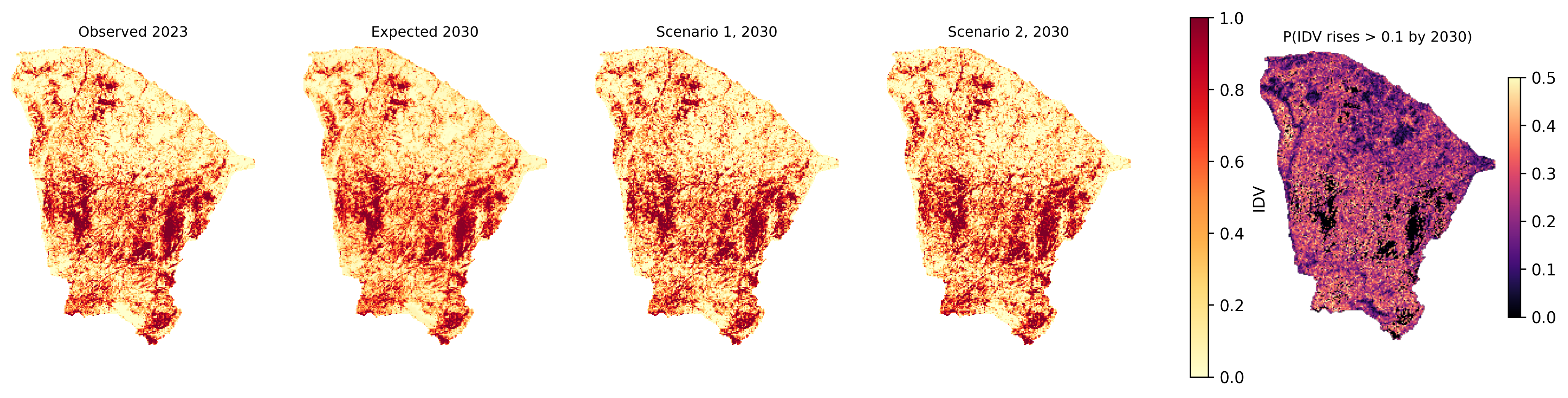}
\caption{From left: observed IDV in 2023, expected ConvLSTM projection for 2030, two stochastic scenarios for 2030, and the fraction of 40 scenarios in which a cell's IDV rises by more than 0.1 between 2023 and 2030.}
\label{fig:scenario_maps}
\end{figure*}

Fig.~\ref{fig:scenario_maps} shows two scenarios for 2030 next to the expected map and the probability that a cell's IDV rises by more than 0.1 by 2030. The scenarios keep the stable areas and the hotspots of the observed maps. In 2030 the share of cells at zero is 0.186, close to the 0.193 of 2023, and the share of saturated cells grows from 0.039 to 0.053, which the hindcast suggests is somewhat too fast. The fraction of the state above the degradation threshold (the 70th percentile of 2000--2023) rises from 36.0\% in 2023 to 38.3\%, against 37.2\% in the expected forecast. The state-wide mean of the scenarios in 2030 is 0.371 (5--95\% range 0.369--0.372), the same as the expected projection. Individual scenarios are plausible realizations, not predictions of where change will happen. The probability map is more useful for planning: it is low on the existing hotspots, which can hardly degrade further, and on the stable vegetated areas of the north, and highest across the intermediate, partly degraded land of the central and southern sertão.

\subsection{Physical Consistency}
\label{sec:consistency_results}

The projection passes all five checks. Four of them are computed over the observed and projected series together, 2000--2030, and none of them shows a break at the transition from observation to projection. Over that period the degraded-area fraction changes by at most 1.2 percentage points from one year to the next (limit 15), the hotspots of consecutive years overlap with a mean IoU of 0.91 (limit 0.2), the centroid of the degraded area moves by at most 1.2 cells (limit 30), and the lag-one autocorrelation of the mean IDV is 0.997 (limit 0.4). The fifth check looks at the projection alone. Its largest annual expansion of the degraded area is 0.25 percentage points, a fifth of the largest observed between 2000 and 2023 (1.2 points). As noted in Section~\ref{sec:consistency_methods}, these checks only show that the projection does not move or inflate degradation implausibly. A model close to persistence passes them almost by construction, so they cannot distinguish a good forecast from a cautious one.

\subsection{Uncertainty}
\label{sec:uncertainty_results}

\begin{figure}[!t]
\centering
\includegraphics[width=\columnwidth]{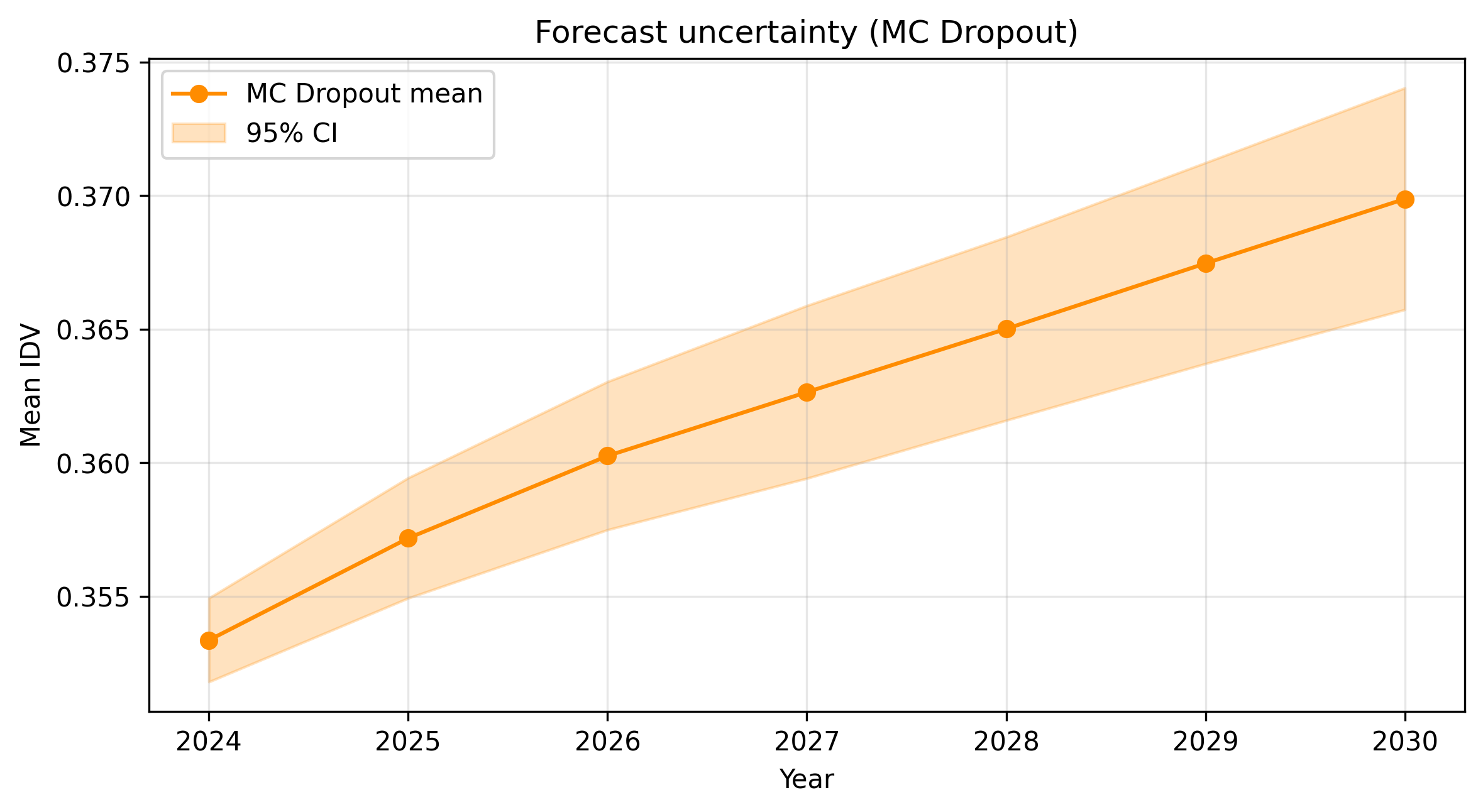}
\caption{State-wide mean of the MC Dropout projection, 2024--2030. The band is the average over valid cells of the cell-level 95\% intervals.}
\label{fig:uncertainty_mc_dropout}
\end{figure}

\begin{table}[!t]
\renewcommand{\arraystretch}{1.2}
\caption{Mean MC Dropout Standard Deviation over Valid Cells, by Forecast Year}
\label{tab:uncertainty}
\centering
\footnotesize
\setlength{\tabcolsep}{2.5pt}
\begin{tabular}{lccccccc}
\toprule
\textbf{Estimator} & 2024 & 2025 & 2026 & 2027 & 2028 & 2029 & 2030 \\
\midrule
MC Dropout ($D=50$) & .0008 & .0012 & .0014 & .0017 & .0018 & .0019 & .0021 \\
\bottomrule
\end{tabular}
\end{table}

Table~\ref{tab:uncertainty} and Fig.~\ref{fig:uncertainty_mc_dropout} summarize the MC Dropout spread, which grows steadily with lead time, from 0.0008 in 2024 to 0.0021 in 2030, as the stochastic rollouts drift apart. Averaged over the state, the 95\% band in 2030 runs from 0.366 to 0.374, entirely above the 2023 value, so the sign of the projected change is robust to the model's own uncertainty. These spreads are about 25 times smaller than the year-to-year change of a typical cell. It is a spread of the model's own predictions and does not account for the cell-level variability that Section~\ref{sec:predictability_results} shows to be unpredictable, which has a standard deviation of about 0.05 per year. A forecast for a single cell is uncertain by at least that amount whatever the reported spread, and the reported spreads are better read as the uncertainty of the projected regional trend.

\section{Discussion}

Desertification is usually described as a cumulative process, the product of long-term interactions among vegetation degradation, climatic variability, hydrological processes, land surface conditions, and other environmental factors.\cite{abdelrahman2023overview,dacunto2024rural,dodorico2013global,li2021drivers}. Our results agree with that description at the scale of the region and disagree with any expectation of predictability at the scale of the pixel. The degradation index in Cear\'a has risen steadily for 24 years, and that rise is visible and partly forecastable. What happens to an individual 2.2~km cell in a given year is mostly not. The median annual change is zero in every IDV class, the typical change is ten times larger than the trend, and a flexible model with access to every channel explains about 5\% of it.

Part of this unpredictability comes from the target. The IDV is derived from annual land-cover maps, and year-to-year reclassification between pasture, other non-vegetated land and natural cover produces jumps that are partly real land-use change and partly classification noise. Neither is well predicted by the climate and vegetation of the preceding years. The other part comes from what EO can see. Every variable used here describes the surface. None directly observes groundwater, soil processes, land tenure, grazing pressure or the economic decisions that lead a farmer to clear or abandon a plot. The satellite record captures the surface expression of degradation, and the gap between that expression and its causes limits forecasting no matter which architecture is used. The gradient-boosting reference turns this argument into a number for Cear\'a: with the information in these ten products, the predictable share of annual cell-level change is close to 5\%.

This has a direct consequence for how forecasting models should be evaluated. Persistence reaches $R^2$ and SSIM values close to one in this problem because the spatial pattern of degradation changes slowly. Any model that copies the last map scores almost as well, so high values of these metrics say little about whether a model has learned dynamics. The earlier sigmoid model is a clear example. Its maps looked right and its scores were close to those of persistence, but its predicted changes were uncorrelated with the observed ones, and when its output was fed back over seven years those errors accumulated into a decline in degradation that the data do not support. Errors in a recursive forecast do not cancel. They are fed back into the input, and when they depend on the state, as reconstruction errors near the bounds of $[0,1]$ do, they push the whole field in one direction. A residual output that starts from persistence makes the learned part of the forecast explicit, and correlating that part with the observed change is a direct test of skill. We think this test, together with drift-aware baselines, should be standard when deep learning is applied to persistent EO fields.

The regional pattern of the drift also matters. The south of the state degraded about twice as fast as the north over 2000--2023, and both regions accelerated after 2017. A model trained on the north and tested on the south is therefore asked to extrapolate to faster change than it has seen, and even a correct drift estimate from the training region falls short there, as the drift baselines show. Similar spatial heterogeneity in the rate of degradation has been reported across drylands, where degradation tends to persist and intensify in places already under pressure \cite{dodorico2013global,ibrahim2015land,li2021drivers}. The channel knockouts point in the same direction. No channel changes the error by more than 1\%. Land surface temperature is the one the network relies on, which fits its role as a direct measure of the surface energy balance and of vegetation water stress. The vegetation indices, often the most informative observable proxies of degradation because they integrate water availability, thermal stress and land management \cite{barbosa2019assessment,higginbottom2014assessing}, and elevation slightly degrade the forecast in the validation band. Much of what the vegetation indices carry about the state of the land is already in the last IDV map, which the residual model uses directly, and what the network learned from them and from elevation in the north appears to be specific to the training region.

The practical message is that EO-based forecasting of desertification is most useful at the scale where it is predictable. The regional trend, and its uncertainty, can be projected with some confidence from the record. Maps of individual cells are mostly the present pattern carried forward, and presenting them as predictions of where degradation will appear next would overstate what the data support. Because every variable comes from globally available products \cite{gorelick2017google,mapbiomas2024}, the same predictability analysis can be repeated in other drylands at low cost, and comparing the predictable share of change across regions would be a useful way to see how general this limit is.

\section{Limitations and Ongoing Work}
\label{sec:limitations}

All results come from forecasting under partial observability. Groundwater, soil degradation, land management and socioeconomic pressures are at best indirectly observed in the products used here, and our predictability estimate is conditional on these ten variables at 2.2~km and annual resolution. Richer inputs could raise it.

The record is short. Twenty-four years limit both the training sample and the statistical power of every comparison. The validation set is also small: 18 patch-windows from a single band in the far south of the state, which makes the validation metrics sensitive to the particular trajectory of that region in 2018--2023. The 64 rows between the training and validation blocks, about a quarter of the grid, are not used at all. A blocked cross-validation over several bands, with the middle of the state included, would give more stable estimates.

The zone-stratified evaluation does not retrain the model with zones withheld, and the channel knockout does not retrain without the removed channels. Both measure properties of one trained network. A strict zone holdout and retrained ablations would answer the questions of transfer and information content more directly.

Some data choices affect the meaning of the variables. All channels were aggregated to 1~km with a maximum-value reduction. For the IDV this makes a cell degraded if any part of it is, and for continuous variables such as NDVI and LST it biases values upward. A mean reduction for continuous channels should be compared. SPI-12 was standardized with a gamma fit over the full 2000--2023 record, so it contains a small amount of information from the validation years. MapBiomas classification noise is part of the target, and separating it from real land-cover change would sharpen the predictability estimate.

The 2024--2030 projection uses climatological forcing. It does not account for future droughts, El Ni\~no events or changes in land use, and its MC Dropout bands reflect model spread, not the cell-level variability described in Section~\ref{sec:predictability_results}. The scenarios add that variability, but they assume that the jump statistics of 2000--2023 continue unchanged.

The benchmark is limited to persistence, drift baselines, a CNN, an LSTM and the ConvLSTM. Given how little of the change is predictable, we do not expect larger architectures alone to help much. Better targets and better inputs seem more promising: forecasting the IDV aggregated to coarser units or regions, probabilistic models of transitions between land-cover classes, explicit treatment of the regional drift, additional observations related to land use and water, and representations from multimodal EO foundation models.

Cear\'a was chosen as a representative semi-arid testbed. Whether the same limits apply in other drylands is still open, and repeating the analysis elsewhere would separate region-specific dynamics from limits that come from partial observability in general.

Several of these points are being addressed in the version of this work under preparation. First, we are retraining the network without the channels that degraded the knockout results (elevation and EVI), and with the other terrain and climate groups removed in turn, so that the information content of each group is measured on models trained without it and not only on one trained network. Second, we are replacing the single northern/southern split with a blocked cross-validation over several latitudinal bands, including the middle of the state, which should also supply the strict zone holdout missing here. Third, we are widening the grid of shrinkage factors applied to the predicted change, since the optimum reached the upper edge of the current grid, and comparing mean and maximum aggregation for the continuous channels. Fourth, we are building conditional projections in which the climatological forcing is replaced by wet and dry sequences, including El Ni\~no and La Ni\~na years, to test how much of the regional drift depends on the climate state. Finally, we are looking for additional data, both longer and finer records for Cear\'a and datasets from other semi-arid regions, so that the same predictability analysis can be applied to other drylands and show whether the 5\% ceiling is specific to Cear\'a.

\section{Conclusion}

We used a ConvLSTM trained on ten EO variables over 2000--2023 to probe how predictable desertification dynamics are in Cear\'a. Measured directly from the data, the mean degradation index rose by about 0.005 per year, faster in the south, while the year-to-year change of individual 2.2~km cells was mostly unpredictable. A flexible model given every channel explained about 5\% of it, and no simple predictor improved on persistence one year ahead at the state scale. A residual ConvLSTM that starts from persistence and learns only the change improved on persistence by about 1--2\% one year ahead, in every validation year, and by about 4\% four years ahead, and it reproduced a little over half of the regional drift. That is close to the ceiling set by the data in the south of the state and well short of what a perfect forecast of individual cells would need. An earlier model that redrew the whole field each year matched persistence on standard metrics while predicting changes unrelated to the observed ones, and its recursive projection showed a spurious decline in degradation. Under climatological forcing, the projection for 2024--2030 continues the rise of the mean IDV, from 0.350 to 0.370, at about half the recent rate, so it is more likely to underestimate degradation in 2030 than to overstate it.

The main conclusion is that, with current EO products, desertification can be forecast as a regional trend more than as a map of where degradation will appear next. Evaluating deep learning models against persistence, drift baselines and the correlation of predicted and observed change makes this limit visible, and the same analysis can be applied to any dryland covered by global EO archives.

\section*{Acknowledgment}
The authors express their sincere gratitude to Matteo Tortora, University of Genoa, for his generous advice on this work and for his interest in its direction. His encouragement and willingness to share his experience have been very valuable to us.

\end{document}